\documentclass[aps,pre,twocolumn,floatfix,a4paper,nofootinbib,amssymb,amsmath,superscriptaddress]{revtex4}
\usepackage{graphicx}
\usepackage{bm}
\usepackage{eucal}
\usepackage{mathrsfs}

\newcommand {\be}{\begin{equation}}
\newcommand {\ee}{\end{equation}}

\usepackage{xcolor}

\begin{document}

\title{Chimera states in coupled Kuramoto oscillators with inertia}
\date{\today}

\author{Simona Olmi}
\email{simona.olmi@fi.isc.cnr.it}
\affiliation{CNR - Consiglio Nazionale delle Ricerche - Istituto dei Sistemi 
Complessi, via Madonna del Piano 10, I-50019 Sesto Fiorentino, Italy}
\affiliation{INFN Sez. Firenze, via Sansone, 1 - I-50019 Sesto Fiorentino, Italy}

\begin{abstract}
The dynamics of two symmetrically coupled populations of rotators is studied for different values of the inertia. The system is characterized by different types of solutions, which all coexist with the fully synchronized state. At small inertia the system is no more chaotic and one observes mainly quasi-periodic chimeras, while the usual (stationary) chimera state is not anymore observable.
At large inertia one observes two different kind of chaotic solutions with broken symmetry: 
the {\it intermittent chaotic chimera}, characterized by a synchronized population 
and a population displaying a turbulent behaviour, and 
a second state where the two populations are both chaotic but
whose dynamics adhere to two different macroscopic attractors. 
The intermittent chaotic chimeras are characterized by a finite life-time, whose duration 
increases as a power-law with the system size and the inertia value. 
Moreover, the chaotic population exhibits 
clear intermittent behavior, displaying a laminar phase where the two populations tend to synchronize, 
and a turbulent phase where the macroscopic motion of one population is definitely erratic.
In the thermodynamic limit these states survive for infinite time and the laminar regimes tends to disappear, thus giving rise to stationary chaotic solutions with broken symmetry contrary to
what observed for chaotic chimeras on a ring geometry.
\end{abstract}


\maketitle
In 2002, simulations of abstract mathematical models revealed the existence of counterintuitive ``chimera states'',
where an oscillator population splits into two parts, with one synchronizing and the other oscillating incoherently, 
even though the oscillators are identical. Since then, these counterintuitive states have become a relevant subject 
of investigation for experimental and theoretical scientists active in different fields, as testified by the rapidly 
increasing number of publications in recent years (for a review see \cite{motter2010, panaggio2015}). 
In this paper we analyze novel chimera states emerging in two symmetrically coupled populations of oscillators with inertia.
In particular, the introduction of inertia allows the oscillators to synchronize via the adaptation 
of their own frequencies, in analogy with the mechanism observed in the firefly \textit{Pteroptix
malaccae}~\cite{ermentrout1991}. The modification of the classical Kuramoto model with the 
addition of an inertial term results in first order synchronization transitions and complex hysteretic 
phenomena~\cite{tanaka1997first,tanaka1997self, Ji2013, gupta2014, komarov2014, olmi2014}.
Furthermore, networks of rotators have recently found applications in
different technological contexts, including disordered arrays of Josephson 
junctions~\cite{trees2005} and electrical power grids~\cite{salam1984,filatrella2008,rohden2012, motter2013}
and they could also be relevant for micro-electro-mechanical systems and optomechanical crystals, where chimeras 
and other partially disordered states likely play an important role with far reaching ramifications.

\section{Introduction}
Collective synchronization is an ubiquitous phenomenon that pervades nature at every scale, and 
underlies essential processes of life; it is a central process observed in a spectacular range of systems, 
such as pendulum clocks~\cite{Huygens1967ab}, pedestrians on a bridge locking their gait~\cite{Strogatz2005}, 
Josephson junctions~\cite{Wiesenfeld1998},  the beating of the heart~\cite{Michaels1987}, 
circadian clocks in the brain~\cite{Liu1997}, chemical oscillations~\cite{Kiss2002}, metabolic oscillations in 
yeast~\cite{Dano1999}, life cycles of phytoplankton~\cite{Massie2010}. In particular,
synchronized oscillations have received particular attention in neuroscience 
because they are prominent in the cortex of the awake brain
during attention and are believed to be involved in higher level processes, such as
sensory binding, awareness, memory storage and replay, and even consciousness~\cite{buzsaki2006rhythms}.
From the clinical point of view, abnormal synchronization seems to play a
crucial role in neural disorders such as Parkinson, epilepsy and essential
tremor~\cite{uhlhaas2006}.

About a decade ago, a peculiar state was theoretically revealed~\cite{Kuramoto2002},
where a population of identical coupled oscillators can split up into two parts where one part synchronizes 
and the other oscillates incoherently. This so-called chimera state is counter-intuitive as it appears even 
when the oscillators are identical, but, since its discover, this state has become
a relevant subject of investigation for experimental and theoretical scientists active in different fields
ranging from laser dynamics to chemical oscillators, from mechanical pendula to (computational)
neuroscience. In particular chimera states have  been shown to emerge in various numerical/theoretical studies
~\cite{abrams2004,Shima2004,abrams2008,Martens2010bistable,Omelchenko2008, Pikovsky2008,Olmi2010,
Laing2009,laing2009physicaD,bastidas2015} and in various experimental settings, 
including mechanical~\cite{MartensThutupalli2013, kapitaniak2014, olmi2015}, 
(electro-)chemical ~\cite{Tinsley2012,Wickramasinghe2013,schmidt2014} lasing systems~\cite{Hagerstrom2012,larger2013} 
and BOLD fMRI signals detection during resting state activity~\cite{cabral2011}, among others. Therefore chimera 
states are an ubiquitous phenomenon in nature much like synchronization itself and may often have been overlooked 
or dismissed in previous studies.

A categorization of different behaviors shown by incoherent oscillators in chimera states has seen 
the emergence of almost regular macroscopic dynamics which are either stationary,
periodic (so-called breathing chimera) or even quasi-periodic~\cite{abrams2008,Pikovsky2008}.  
Only recently, spatio-temporally chaotic chimeras have been numerically identified
in coupled oscillators on ring networks~\cite{bordyugov2010,wolfrum2011,omelchenko2011loss, sethia2014}
and in globally connected populations of pulse-coupled oscillators~\cite{pazo2014}.
However, a detailed characterization of the dynamical properties of these states have been reported only 
for the former case, more specifically, for rings of nonlocally coupled phase oscillators: in this case chimeras 
are transient, and {\it weakly chaotic}~\cite{wolfrum2011,wolfrum2011spectral}. In particular, the life-times of 
these states diverge exponentially with the system size, while their dynamics becomes regular in the thermodynamic limit. 

In a recent paper \cite{olmi2015} it has been shown that a simple model of coupled oscillators with inertia 
(rotators) can reproduce dynamical behaviours found experimentally for two coupled populations of mechanical pendula. 
Starting from these results, the  present paper is devoted to a detailed analysis of the collective states emerging in
such model of symmetrically coupled rotators, for different 
inertia values and sizes. For small inertia values breathing and quasiperiodic chimeras coexist with the synchronized state, 
while two chaotic solutions emerge for sufficiently large inertia: the
chaotic chimera, characterized by a synchronized and a chaotic population, and a state where both
populations are chaotic, but with two distinct macroscopic attractors. While the last chimeras are stationary states,
the former ones are characterized by a finite life-time, whose duration diverges as a power-law with the system size and
the inertia.  On one hand the chaotic population exhibits clear intermittent behavior between a laminar and a
turbulent phase; in particular, in the turbulent regime, the Lyapunov analysis reveals that the stability properties
of chaotic chimeras can be ascribed to the universality class of globally coupled systems~\cite{takeuchi2011}.
On the other hand, in the two chaotic population states, the most part of neurons belong to an unique cluster 
and the chaotic evolution is driven only by the oscillators out of the cluster.

Moreover, a numerical extension of the zero-inertia solution has been performed, starting from a stationary chimera state, 
characterized by a synchronized and a partially synchronized population, where both order parameters are constant. 
Due to the introduction of the inertia, the order parameter of the partially synchronized population is no longer constant 
but it oscillates periodically about the zero-inertia limit value. Therefore, in presence of inertia, stationary chimeras are 
any longer observed; only breathing and quasiperiodic chimeras emerge.
In particular in Sec. II we will introduce the model, the order parameters employed to characterize the level of coherence
in the system, and we will describe our simulation protocols as well as the linearized system. In Sec. III we will analyze the 
different stationary states emerging in the system for different inertia values and different initial conditions.
In Sec. IV we will report the stability properties of the intermittent chaotic chimera emerging at 
sufficiently large inertia value. In Sec. V the microscopic dynamics of the two chaotic population state is deeply 
investigated. Finally, the reported results are briefly summarized and discussed in Sec. VI.
The finite size scaling of the maximal Lyapunov exponent is reported in Appendix A, while in the Appendix B the linear stability of 
the synchronized state is presented.

\section{Model and Tools}

\subsection{Model and Macroscopic Indicators}
We consider a network of two symmetrically coupled populations of $N$ rotators,
each characterized by a phase $\theta_i^{(\sigma)}$ and a frequency
$\omega_i^{(\sigma)} \equiv {\dot \theta}_i^{(\sigma)}$, where $\sigma=1,2$
denotes the population. The phase $\theta_i^{(\sigma)}$ of the $i$-th oscillator 
in population $\sigma$ evolves according to the differential equation
\begin{equation}
\label{eq1} 
m\ddot{\theta}_i^{(\sigma)} + \dot{\theta}_i^{(\sigma)}=\Omega+\sum_{\sigma'=1}^2 
\frac{K_{\sigma\sigma'}}{N} \sum_{j=1}^N  \sin{\left(\theta_j^{(\sigma')}-\theta_i^{(\sigma)}-\gamma\right)}\,,
\end{equation}
where the oscillators are assumed to be identical with inertia $m$, 
natural frequency $\Omega=1$ and a fixed phase lag $\gamma= \pi - 0.02$. 
The self- (cross-) coupling among oscillators belonging to the same population 
(to different populations) is defined as $K_{\sigma\sigma} \equiv \mu$ ($K_{\sigma\sigma'} =K_{\sigma'\sigma} \equiv \nu$),
with $\mu+\nu=1/2$ without loss of generality.
We follow previous studies on chimera states~\cite{abrams2008,montbrio2004} and impose an imbalance between intra- 
and inter-population interactions quantified by $A = (\mu-\nu)/(\mu+\nu)$ with $\mu>\nu$. 
Thus, uniform coupling is achieved when $A=0$, whereas the populations are disconnected for $A=1$. 
In the following analysis $A$ will be kept equal to 0.2.

We  consider only two types of initial conditions: uniform (UCs) or with broken symmetry (BSCs). 
In the former case both populations are initialized with random values; in the last case the populations
are initialized differently: one population is initialized in a fully synchronized state and it
has a set of identical initial values for both phases and frequencies
(namely, ${\theta}_i^{(\sigma)} = \dot{\theta}_i^{(\sigma)} \equiv 0$ $\forall i$), 
while the phases and frequencies of the second population can be initialized either with 
random values or with equispaced values taken from the intervals  ${\theta}_i^{(\sigma')} \in [-\pi:\pi]$ 
and $\dot{\theta}_i^{(\sigma')} \in [-\Omega:\Omega]$.
The BSCs can lead to the emergence of Intermittent Chaotic Chimeras (ICCs), a broken symmetry state
where one population is fully synchronized and the other behaves chaotically. In particular this state
exhibits turbulent phases interrupted by laminar regimes. On the other hand the UCs 
can lead to the emergence of Chaotic Two Populations States (C2P), broken symmetry states where both populations 
display chaotic behavior taking place on different attractors.
The collective evolution of each population will be characterized in terms of the
macroscopic fields 
\begin{equation}
 \rho^{(\sigma)}(t)=R^{(\sigma)}(t) e^{i \Psi(t)} = N^{-1}\sum_{j=1}^N 
e^{i\theta^{(\sigma)}_j(t)}. 
\end{equation}
The modulus $R^{(\sigma)}$ is an order parameter 
for the synchronization transition being one (${\cal O}(N^{-1/2})$) for synchronous 
(asynchronous) states.

In general we will perform sequences of simulations by varying adiabatically
the inertia value $m$ with two different protocols.
Namely, for the first protocol, the series of simulations is initialized from zero-inertia solutions corresponding 
to stationary chimera states for two coupled population of Kuramoto oscillators \cite{abrams2008}.
Starting from these initial conditions, the solutions
are continued to a finite inertia value by employing Eq. (\ref{eq1}), by considering $m_L=1*10^{-4}$ as minimal inertia. 
Afterwards the inertia is increased in small steps $\Delta m$ until a maximal inertia value $m_M=0.001$ is reached. 
For each value of $m$, apart the very first one, the simulations are initialized by 
employing the last configuration of the previous  simulation in the sequence.

For the second protocol, starting from a high inertia value $m=15$, the inertia is reduced until the minimal
non zero inertia value $m_L$ is recovered \footnote{In particular, starting from $m=15$, inertia is decreased to $m=1$ in steps of $\Delta m=1$; from $m=1$ to $m=0.1$ 
the step size is $\Delta m=0.1$; from $m=0.1$ to $m=0.01$ the step size is $\Delta m=0.01$;
from $m=0.01$ to $m=0.001$ the step size is $\Delta m=0.001$; and finally from $m=0.001$ to $m=m_L$ the step size which has been employed is $\Delta m=m_L$.}. The first protocol has been used to investigate the continuation 
to finite inertia of the zero-inertia solution, while the second one has been employed to investigate the emergence 
of different states in the system. At each step the system is simulated for a transient time $T_R$ 
followed by a period $T_W$ during which the average value of the order parameters $\bar{R}^{(\sigma)}$ and  of the 
frequencies $\left\lbrace \bar{\omega}_i^{(\sigma)} \right\rbrace 
\equiv\{\bar{\dot \theta}_i^{(\sigma)} \}$, are estimated.  

\subsection{Lyapunov Analyses}
\label{sectionLyapunov}

The stability of Eq.~(\ref{eq1}) can be analyzed by following the evolution of infinitesimal
perturbations in the tangent space, whose dynamics is ruled by the linearization of
Eq.~(\ref{eq1}) as follows:
\begin{eqnarray}
\label{eq2}
 && m \enskip\delta\ddot{\theta}_i^{(\sigma)} + \delta\dot{\theta}_i^{(\sigma)}= 
 \\ \nonumber
 &&\sum_{\sigma'=1}^2 \frac{K_{\sigma\sigma'}}{N} \sum_{j=1}^N  \cos{\left(\theta_j^{(\sigma')}-
\theta_i^{(\sigma)}-\gamma\right)(\delta\theta_j^{(\sigma')}-\delta\theta_i^{(\sigma)})} \enskip .
\end{eqnarray}

The exponential growth rates of the infinitesimal perturbations are measured in term of the associated
Lyapunov spectrum $\{ \lambda_k \}$, with $k=1, \dots, 4N$. 
In order to numerically estimate the Lyapunov spectrum by employing the method
developed by Benettin {\it et al.}~\cite{benettin1980}, one should consider for each
Lyapunov exponent (LE) $\lambda_k$ the corresponding 4$N$-dimensional 
tangent vector $\mathcal{T}^{(k)} = (\delta\dot{\theta}_1^{(1)},...,\delta\dot{\theta}_N^{(1)},
\delta\dot{\theta}_{1}^{(2)},...,\delta\dot{\theta}_{N}^{(2)},\delta{\theta}_1^{(1)},
...,\delta{\theta}_N^{(1)},\delta{\theta}_{1}^{(2)},...,\delta{\theta}_{N}^{(2)})$,
whose time evolution is given by Eq. (\ref{eq2}).
Furthermore, the orbit and the tangent vectors should be followed for a sufficiently long time lapse $T_s$ by 
performing Gram-Schmidt ortho-normalization at fixed time intervals $\Delta t$, 
after discarding an initial transient evolution $T_t$. 
In the present case we have employed $\Delta t=5$ and  $T_t = 5,000$,
while for BSCs we have integrated the system for times $8 \times 10^4 \le T_s \le 3 \times 10^5$ 
for $N = 100, \dots, 800$ and for UCs for a time range $3 \times 10^4 \le T_s \le 1 \times 10^6$ 
for $N = 100, \dots, 400$. The integrations have been 
performed with a fourth order Runge-Kutta scheme with a time step $5 \times 10^{-4}$.

\begin{figure}
\begin{center}
\includegraphics*[angle=0,width=7cm]{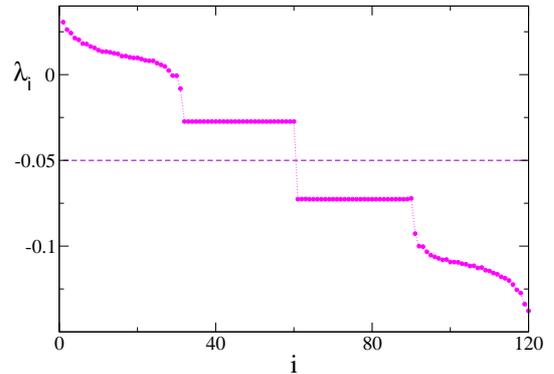}
\end{center}
\caption{ Entire Lyapunov spectrum $ \lambda_i $ 
for an ICC. The symmetry of the spectrum with respect to $-1/2m$ is clearly visible. 
The data refer to $m=10$, $N=30$; the spectrum has been obtained by following the orbit and the
tangent vectors for a time lapse $T_s = 8,000$, after discarding a transient
period $T_t = 1,000$.
}
\label{fig.0}
\end{figure}

It should be noticed that our model differs from a Hamiltonian system just for a constant viscous dissipative 
term proportional to $1/m$, once both sides of Eq. (\ref{eq1}) are rescaled by the inertia of the single oscillator. 
For this class of systems, U. Dressler in 1988 has demonstrated  
that a generalized pairing rule, similar to the one valid for symplectic
systems, applies for the LEs~\cite{dressler1988}, namely
\begin{equation}
\lambda_i+\lambda_{4N-i+1}=-\frac{1}{m} \enskip, \qquad i=1,\dots,2N \enskip.
\label{pairing} 
\end{equation}
Therefore, as shown in Fig. \ref{fig.0} for an ICC state the Lyapunov spectrum is perfectly symmetric 
with respect to $-1/2m$. Due to this property, we estimate only the first part of the spectrum for 
$i=1,\dots,2N$, being the second part obtainable via Eq.~(\ref{pairing}). 

Furthermore, as shown in  \cite{ginelli2011}
the values of the squared components $\xi_i^{(\sigma)}$
of the maximal Lyapunov vector ${\cal T}^{(1)}$ can give important information
about the oscillators that are more actively contributing to the chaotic
dynamics. The squared component for the oscillator $i$  of population $\sigma$ 
is measured as
\begin{equation}
\xi_i^{(\sigma)}(t)=  [\delta\dot{\theta}_i^{(\sigma)}(t)]^2 
+ [\delta{\theta}_i^{(\sigma)}(t)]^2 \quad ;
\label{loc_vect} 
\end{equation}
once the Lyapunov vector is normalized, i.e. $||{\cal T}^{(1)}(t) = 1||$.

Moreover, the occurrence of intermittent laminar and turbulent phases, observable for ICCs, renders the 
characterization of the chaoticity of the system in terms of the asymptotic maximal LE extremely difficult.
Therefore, for such states it is more useful to estimate the  finite time Lyapunov exponents (FTLEs) $\Lambda$  
over a finite time window of duration $\Delta t$, namely
$$ \Lambda =\frac{1}{\Delta t} \ln \sqrt{\sum_{i=1}^{4N} \mathcal{T}_i^{(1)}(\Delta t) 
\mathcal{T}^{(1)}_i(\Delta t)} \enskip;$$ 
where the initial magnitude of the vector is set to one, i.e. $||\mathcal{T}^{(1)}(0)||  \equiv 1$. In particular, 
we measured the associated probability distribution functions $P(\Lambda)$ by collecting $100,000$
data points for each considered system size obtained from ten different orbits, each of duration $T_s = 100,000$
with $\Delta t = 10$. The eventual presence of a peak around $\Lambda \simeq 0$ in the $P(\Lambda)$
indicates the occurrence of laminar phases, usually superimposed over a Gaussian-like profile.
 
In order to give an estimate of the maximal LE, we have removed from the $P(\Lambda)$ the channels eventually 
associated to the laminar phase, and this modified distribution has been fitted with a Gaussian function, 
namely  $F(\Lambda)=\frac{1}{\sqrt{2\pi\sigma^2}} \exp{\left[\frac{(\Lambda-\Lambda^{(*)})^2}{2\sigma^2}\right]}$. 
The maximum of the Gaussian, $\Lambda^{(*)}$ is our best estimate of the maximal LE of the system.

From each $P(\Lambda)$ we have also obtained an estimate of the probability $p_0$ that the chaotic population stays 
in the laminar phase. In particular,  $p_0$ has been measured by integrating
$P(\Lambda)$ over the channels corresponding to $\Lambda=0$ and its nearest-neighbor channels 
(for a total of 3 to 5 channels).

\section{Stationary States}

\begin{figure}[h]
\includegraphics*[angle=0,width=7.cm]{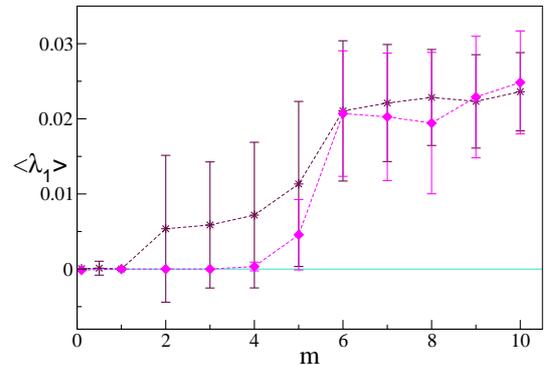}
\caption{Average maximal LE $<\lambda_1>$ and the associated standard deviation vs $m$ for $N=100$: 
magenta diamonds (maroon stars) refer to BSCs (UCs). The  averaged values $<\lambda_1>$ are obtained by following 
each realization for a time span $t=50,000$ and by averaging over 100 different initial conditions. }
\label{fig.diagramma}
\end{figure} 

In this Section we analyze the different stationary states emerging 
in the system for different inertia values and for different initial
conditions: namely, BSCs and UCs.

Starting simulations with BSCs at small inertia ($m \le 4$), 
the system is not chaotic (as shown in Fig.~\ref{fig.diagramma}) and it displays a multitude of coexisting breathing 
and quasiperiodic chimeras~\cite{abrams2008,Pikovsky2008}.
For larger inertia, the system displays not only regular but also chaotic solutions, 
as it can be appreciated by the fact that the maximal Lyapunov exponent, averaged over
many different random realization of the initial conditions, becomes positive (see Fig.~\ref{fig.diagramma}). In particular,  for sufficiently large $m$-values 
ICC solutions with broken symmetry emerges. These solutions, have been previously observed 
experimentally and numerically in \cite{olmi2015} and they will be characterized in 
detail in Sec. V.
 
By considering UCs, the system evolves towards chaotic solutions already at smaller inertia, namely $m > 1$, as shown in Fig.~\ref{fig.diagramma}. With these initial conditions the 
multistability is enhanced and many different coexisting states with broken symmetry 
are observable, either regular or chaotic. The chaotic ones will be examined 
in Sec. VI.

\subsection{Broken Symmetry Initial Conditions} 

\begin{figure} 
\includegraphics*[angle=0,width=7.5cm]{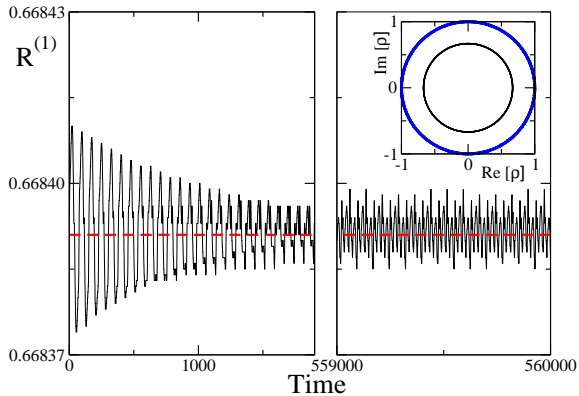}
\caption{Order parameter $R^{(1)}$ for the non-synchronized population 
versus time. The red dashed curve represents the zero-inertia solution $R^{(1)}$
associated to a stationary chimera for $A=0.2$, while the black
curve is $R^{(1)}$ for $m=1 \times 10^{-4}$ continued from the zero-inertia 
state. The two panels report the evolution of $R^{(1)}$ for two successive 
time windows. The inset displays the real and imaginary part of the non-synchronized 
field $\rho^{(1)}$ corresponding to the second panel (black inner curve), while the blue dotted curve refers to 
$\rho^{(2)}$ associated to the fully synchronized population.
The system has been integrated for a time $T_s = 5.6 \times 10^5$ 
with a time step $2 \times 10^{-4}$ and $N=200$ oscillators.
}
\label{fig.prolungamento}
\end{figure}

\subsubsection{Breathing versus Quasi-Periodic Chimeras} 

\begin{figure}
\includegraphics*[angle=0,width=8cm]{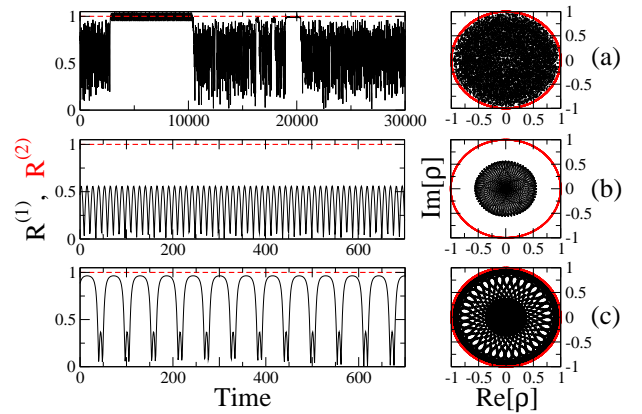}
\caption{Left column: Order parameters $R^{(1)}(t)$ (solid black curve)
and $R^{(2)}(t)$ (dashed red curve) for the two coupled populations versus
time: (a) $m=10$ (chaotic chimera); (b) $m=3$ (breathing chimera); (c) $m=m_L$ (quasiperiodic chimera). 
Right column: real and imaginary parts of the fields $\rho^{(\sigma)}$ corresponding to the different 
dynamical states reported in the left panels. In all cases an initial transient time  $T_t=25$ is discarded. 
For panels (a) and (b) the integration step is $5 \times 10^{-4}$, while for panel (c) 
is $2 \times 10^{-4}$.
The systems are simulated for a time $T_s=5 \times 10^5$ and $N=200$ oscillators.}
\label{fig.staticc}
\end{figure}

At zero inertia, for the considered set of parameters, a stationary chimera
with a constant order parameter $R^{(1)} < R^{(2)} \equiv 1$ has been 
observed by Abrams \textit{et al.} in \cite{abrams2008}. 
In order to verify the stability of the stationary chimera for finite inertia,
we continued the zero-inertia solution to a sufficiently small inertia value (namely, $m_L$). 
In Fig.~\ref{fig.prolungamento} the results of such simulation are reported, in
particular the order parameter for the non-synchronized population is displayed.
Due to the introduction of inertia, the order parameter is no longer constant, but it 
shows initally damped oscillations of extremely small amplitudes (namely, $\simeq 10^{-4}$)
around the zero-inertia value. After a sufficiently long integration time, we observe that
the oscillations stop to decrease and instead reveal a tendence to increase,
but over very long times scales:
in a time window $T=5*10^5$
the amplitude of the fluctuations grows of a factor 10, namely from from $6*10^{-6}$ to $1.6*10^{-5}$.
As a matter of fact, after a simulation time
as long as $T_s=5.6*10^5$, the system has not yet reached a stationary solution. However, for the same inertia 
value $m_L$ we obtain a completely different chimera state
by following the second protocol. This solution is characterized by wide periodic oscillations 
of the order parameter between zero and one, as shown in Fig.~\ref{fig.staticc}(c). 

In order to better understand the difference among these two states we  have
estimated for the non synchronized population the Fourier power spectra $P_s(\nu)$
for the real and imaginary part of the field $\rho^{(1)}$ and for its modulus $R^{(1)}$. 
The state obtained by the continuation of the zero-inertia solution reveals a single peak in the spectra 
of the real (imaginary) part of $\rho^{(1)}$ at a frequency $\nu_1 \simeq 0.091$. Therefore,
according to the definition reported in \cite{abrams2008}, this can be classified as a breathing chimera.
Conversely the Fourier spectrum for the real part of the signal reported in Fig.~\ref{fig.staticc}(c) 
reveals two main peaks at uncommensurable frequencies, one at $\nu_1$ and a new one at $\nu_2 \simeq 0.073$,
while the Fourier spectrum of the order parameter has an unique peak at a frequency $\nu_1-\nu_2$. In this case 
the field is quasi-periodic, while its modulus is  periodic, and this can be easily achieved by setting,
e.g. $Re[\rho]= A \cos(2 \pi \nu_1 t) + B \cos(2 \pi \nu_2 t)$
and $Im[\rho]= A \sin(2 \pi \nu_1 t) + B \sin(2 \pi \nu_2 t)$.
This is the only type of chimera state we have found in our simulations with BSCs by employing
the second protocol and for all the considered inertia values. These can be classified as 
quasi-periodic chimera, according to \cite{Pikovsky2008}. 
Due to our limited CPU resources, we cannot explore infinite times and 
we cannot exclude that, for sufficiently long times, the solution obtained 
by following the first protocol will reveal the emergence of a 
a second frequency $\nu_2$ and the state will eventually converge towards 
the solution obtained by following the second protocol.

If the solution is continued to inertia values larger than $m_L$, we observe 
wider oscillations around the zero-inertia stationary value. But the
dynamical behaviour is the same: initially we observe damped oscillations, 
which, after a sufficiently long time interval, begin to show a tendency
to increase again. Therefore, we can summarize our results by
affirming that it is not possible to continue exactly the stationary chimeras 
obtained in \cite{abrams2008}. Furthermore, while the fully synchronous state   
remains stable also in presence of inertia, as shown in Sect. III,
the stationary chimeras with constant order parameters are no 
longer observable at $m>0$: only breathing or quasi-periodic chimeras are present.
This is probably due to the fact that the addition of inertia to the
phase oscillators' models corresponds to the addition of a  
degree of freedom to the system, and this is a sort of singular
perturbation. Finally, we have shown that chimera states can exist for inertia
values as small as $m_L = 1 * 10^{-4}$, a value definitely smaller
than $m=1*10^{-3}$, reported as {\it insuperable threshold}
by Bountis \textit{et al.} in~\cite{bountis2014}. We think
that the results reported in~\cite{bountis2014} should be due to
integration inaccuracies and we are confident that chimera states at inertia 
values even smaller than $m_L$ can be observed, it is just a matter of employing 
sufficiently accurate integration schemes.

\subsubsection{Coexisting quasi-periodic chimeras.}

By employing the second protocol with BSCs, and
starting from $m=15$ the inertia is decreased to $m=m_L$ as explained in SubSect. II A.
In the present case we have also performed a refined numerical analysis in the range $5\leq m \leq 6$ to  
determine the critical inertia value $m_c$ at which we observe the 
transition from chaotic to quasi-periodic chimeras, we have found $m_c=5$.
The system shows a multitude of broken symmetry states, peculiar examples are reported
in Fig.~\ref{fig.staticc}. For sufficiently high inertia ($5\leq m<15$) ICCs are always 
observable (see Fig.~\ref{fig.staticc} (a)); these states will be discussed in details in Sect. V.
For lower inertia values we observe only quasi-periodic chimeras of the previously described
type: quasi-periodic in the macroscopic field and periodic in its modulus.
Two examples are show in Fig.~\ref{fig.staticc} (b) and (c) for $m=3$ and $m_L$,
in the first case $R^{(1)}(t)$ reveals an almost sinusoidal shape, while  in the other case not.

\begin{figure}
\includegraphics*[angle=0,width=6cm]{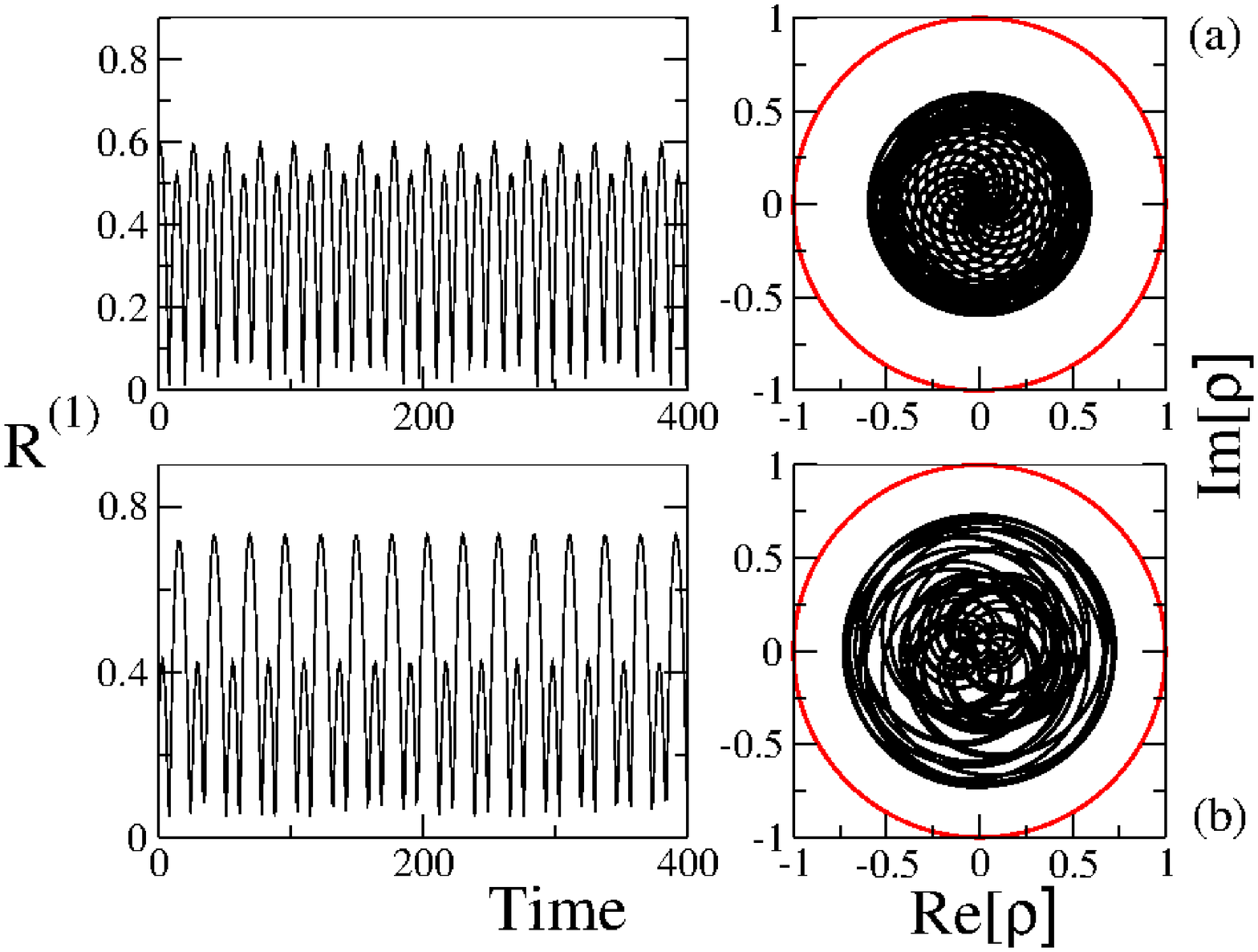}
\includegraphics*[angle=0,width=6cm]{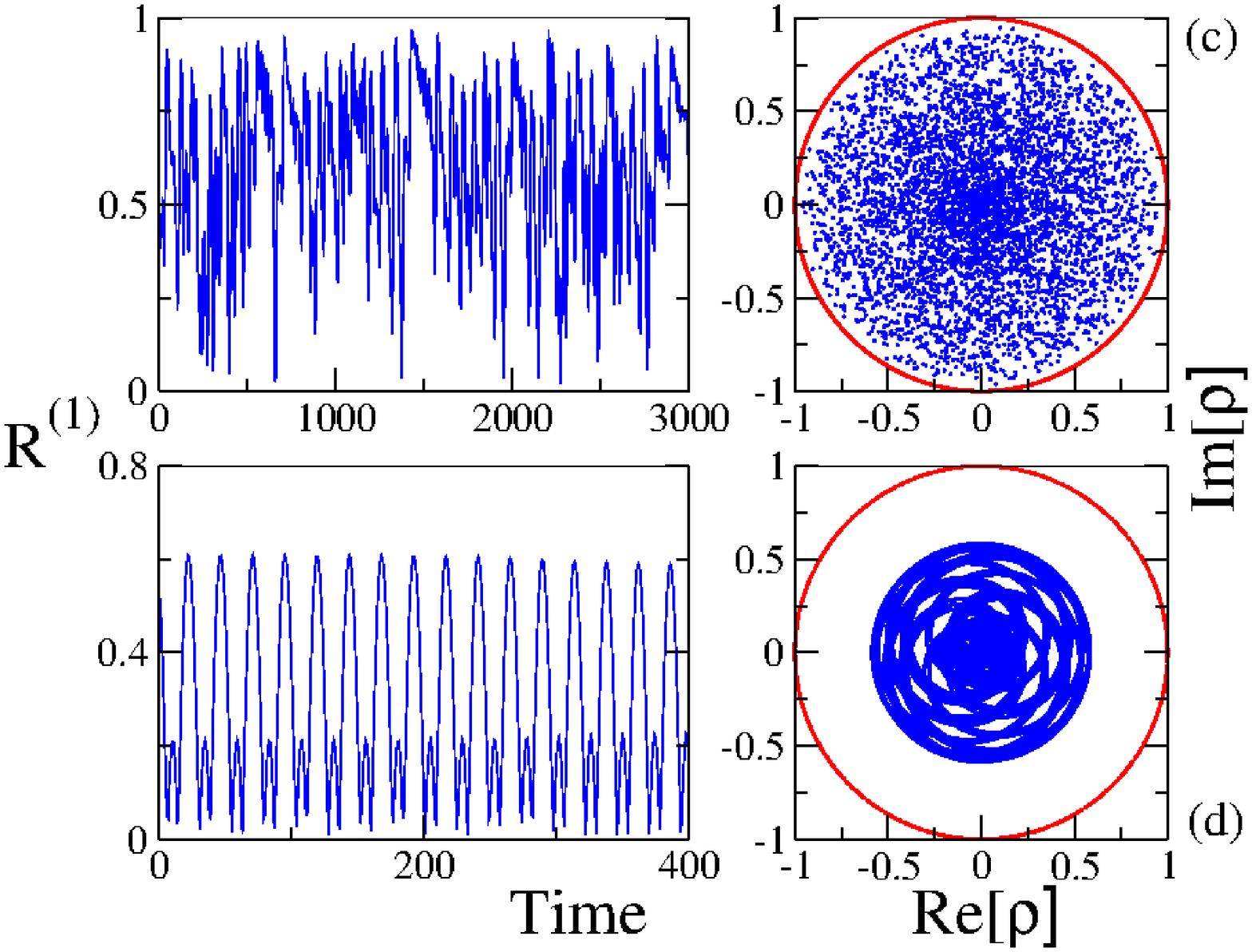}
\caption{Left panels: Order parameter of the non-synchronized family as a function of time for different BSCs. 
The order parameter of the synchronized family is not shown. Right panels: imaginary vs real part of the complex fields
$\rho^{(1)}$ and $\rho^{(2)}$ for $m=3$ (panels (a),(b)) and for $m=5$ (panels (c),(d) ). 
In all cases a time step $5 \times 10^{-4}$ is used and an initial transient time 
$T_t=25$ is discarded.
The systems are simulated for a time $T_s=5 \times 10^5$ and $N=200$ oscillators.}
\label{fig.multistabilityCC}
\end{figure}

Furthermore, the system is multistable, for different initial conditions a multitude
of coexisting quasi-periodic states are observable. A few examples are reported in Fig. ~\ref{fig.multistabilityCC} 
for $m=3$ and 5. At the lower inertia value $m=3 < m_c$ non chaotic quasi-periodic chimeras are present with 
different periods and level of synchronization, while for $m=5$ non chaotic quasi-periodic chimeras coexist
with ICCs. 

\subsection{Uniform Initial Conditions} 

\begin{figure}[h]
\includegraphics*[angle=0,width=4.75cm]{f6a.eps}
\includegraphics*[angle=0,width=3.8cm]{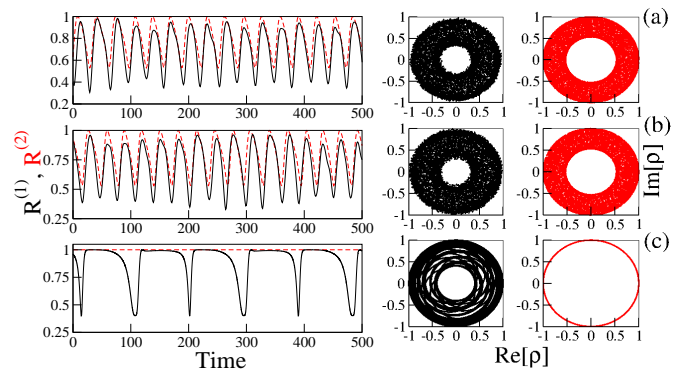}
\caption{Left panels: Order parameters $R^{(1)}$ (black solid line) and $R^{(2)}$ (red dashed line) vs time; Right panels:  imaginary vs real part of the complex fields 
$\rho^{(1)}$ and $\rho^{(2)}$  for (a) $m=10$; (b) $m=9$; (c) $m=3$.
In all cases a time step $5 \times 10^{-4}$ is used and an initial transient time 
$T_t=25$ is discarded. The systems are simulated for a time $T_s=100,000$ and for $N=200$.}
\label{fig.statictp}
\end{figure}

By following the second protocol with UCs, the inertia value is decreased from $m=15$ to $m_L$ as explained in 
SubSect. II A. Also with UCs the system shows a moltitude of broken symmetry states, as shown 
in Fig.~\ref{fig.statictp}. In particular, C2P solutions are always present in the range $5 \leq m \leq 15$. 
These states are characterized by an irregular behavior of both order parameters
and the dynamics of the two populations takes place on different macroscopic attractors, as it is clearly observable 
in Fig.~\ref{fig.statictp} (a) and (b). These states will be discussed in details in Sec. \ref{C2P}.
In the intermediate inertia range $3\leq m< 5$ the system exhibits broken symmetry states, 
where the macroscopic field is quasi-periodic and its modulus is periodic. Even though the macroscopic 
field shows an almost regular behavior, a weak chaoticity is still present (see Fig ~\ref{fig.statictp} (c)). 
This residual chaoticity seems to be related to the distortion in amplitude of the macroscopic field
and can be observed only following the second protocol with UCs. Finally for smaller inertia values $(m_L\leq m<3)$ only the fully synchronous state is present.

\begin{figure}[h]
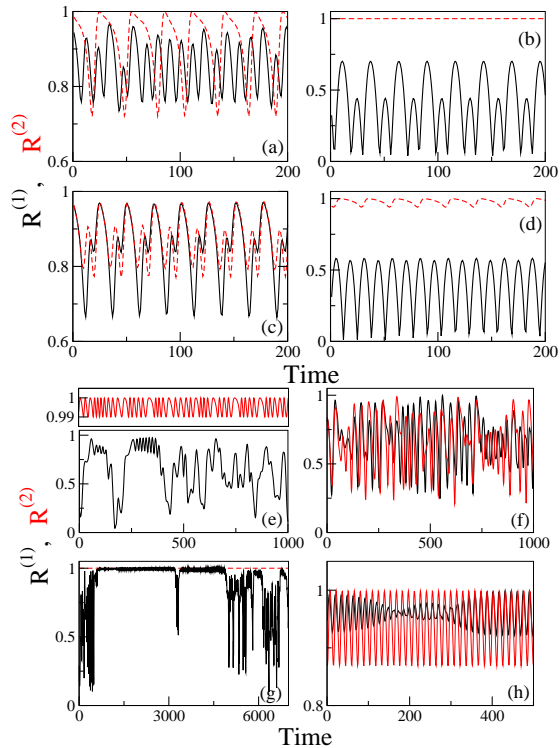

\includegraphics*[angle=0,width=7.2cm]{f7a.eps}
\includegraphics*[angle=0,width=7cm]{f7b.eps}
\caption{Order parameters $R^{(1)}$ (black solid line) and $R^{(2)}$ (red dashed or solid line) vs time for different UCs. 
Panels (a)-(d) refer to $m=3$. Panels (e)-(h) refer to $m=10$. 
The systems are simulated for a time $T_s=100,000$ and for $N=200$.}
\label{fig.multistabilityC2P}
\end{figure}

The system is highly multistable; in particular, if it is initialized at $m=3$ and 2 with UCs, without following any protocol, quasi-periodic chimeras with no residual chaoticity (Fig. ~\ref{fig.multistabilityC2P} (b)) coexist with C2P solutions (Fig. ~\ref{fig.multistabilityC2P} (a)) and with generalized broken symmetry states (Fig. ~\ref{fig.multistabilityC2P} (c), (d)), where both macroscopic fields are quasi-periodic, but the level of synchronicity is different for the populations. Conversely at $m=10$, C2Ps
can be observed (Fig. ~\ref{fig.multistabilityC2P} (f), (h)) as well as ICCs (Fig. ~\ref{fig.multistabilityC2P} (g)) and a 
novel chaotic brooken symmetry state (Fig. ~\ref{fig.multistabilityC2P} (e)), where both population are chaotic, but
one population shows also intermittency. 

\section{Intermittent Chaotic Chimeras}  
\label{chaoticchimeras}

In this Section, we want to better characterize the dynamics of ICCs, observable
with BSCs,  already reported in Fig.~\ref{fig.staticc} (a). 
The chaotic population exhibits clear intermittent behavior (see Fig.~\ref{fig.Frequenze}(a)), 
displaying a laminar phase where the two populations tend to synchronize, and a turbulent phase, 
where the order parameter behaves irregularly. In particular, during laminar phases, 
the non-synchronized order parameter stays in proximity of value one, displaying small oscillations.
Furthermore, as we will show in the following the ICCs are transient, but we have been unable to find with BSCs and for the same inertia coexisting ICCs associated to different chaotic attractors.

\begin{figure}
\includegraphics*[angle=0,width=7cm]{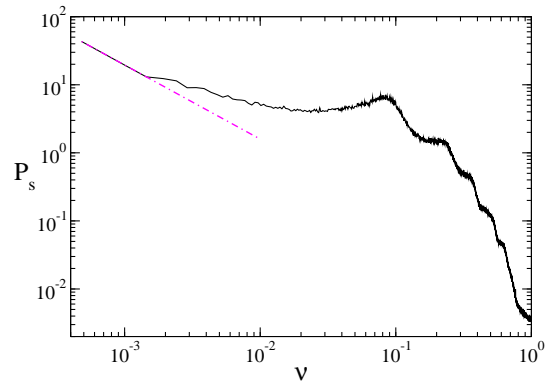}
\caption{Power spectrum $P_s(\nu)$ of the chaotic order parameter as a function of the frequency $\nu = 1/T$ for an ICC. 
Red dashed line: $y = 1.06 \nu^{-1.09}$. Data refer to $m=10$, $N=200$.
The spectrum $P_s$ is obtained by following each realization for a time span $t=100,000$ and by 
averaging over 100 different initial conditions.}
\label{fig.powerspectrumCC}
\end{figure}

A further indication of the presence of intermittency can be given by the estimation of the power spectrum $P_s(\nu)$ of the incoherent order parameter (see Fig. \ref{fig.powerspectrumCC}). We 
observe that $P_s(\nu)$ approaches the limit of low frequencies
as $1/\nu$, as expected for an intermittent system \cite{manneville1980}.
In particular, Manneville in \cite{manneville1980} reported a detailed analysis of a discrete dissipative dynamical system displaying intermittency, revealing that 
a $1/\nu$ spectrum is typically associated to a system exhibiting a transition 
to turbulence via an intermittency route.

\subsection{Finite Time Lyapunov Analysis}

\begin{figure}[h]
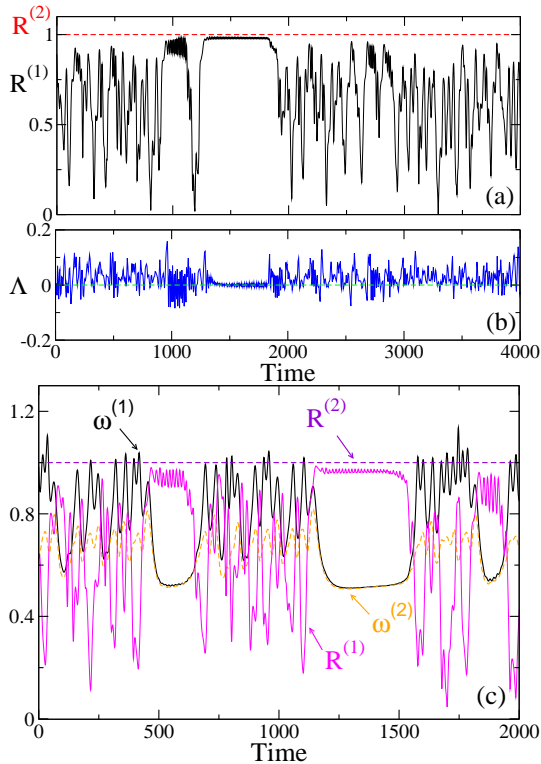

\includegraphics*[angle=0,width=7cm]{f9a.eps}
\includegraphics*[angle=0,width=7cm]{f9b.eps}
\caption{(a) Order parameters $R^{(1)}(t)$ (solid black curve)
and $R^{(2)}(t)$ (dashed red curve) for the ICCs as a function of time. (b) FTLE 
$\Lambda$ for the chaotic population versus time. (c) The solid black (dashed orange) curve represents 
the frequency $\omega^{(1)}$ ($\omega^{(2)}$) of a single oscillator belonging to the chaotic (synchronized) 
population as a function of time; the solid magenta (dashed violet) curve represents its respective order parameter
$R^{(1)}(t)$ ($R^{(2)}(t)$). All simulations are referred to $m=10$, $N=200$.}
\label{fig.Frequenze}
\end{figure}

\begin{figure}
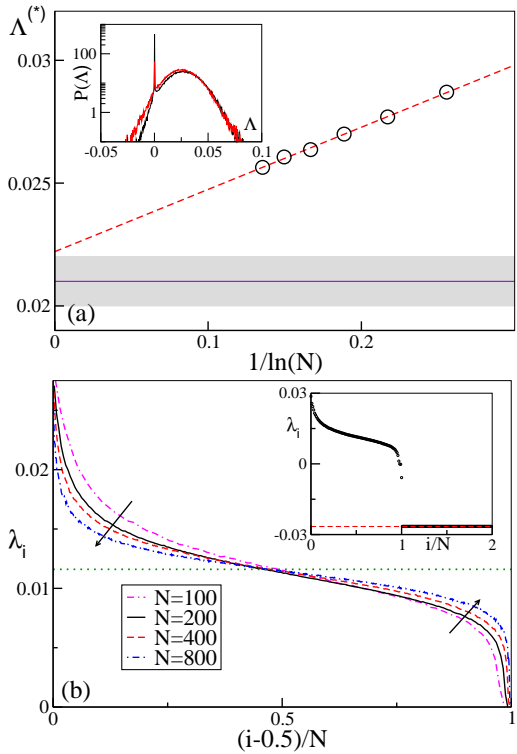

\includegraphics*[angle=0,width=6.7cm]{f10a.eps}
\includegraphics*[angle=0,width=6.7cm]{f10b.eps}
\caption{(a) Plot of $\Lambda^{(*)}$ (symbols) versus $1/ln(N)$ for $100 \leq N\leq 1600$ and the corresponding fit. The continue violet curve with the shaded area denotes $\lambda_M(\infty)$ with its error bar. Inset: Probability distribution function $P(\Lambda)$ for $N=150$ (black curve) and 1600 (red curve). (b) Positive part of the Lyapunov spectra for various sizes, the dotted green line is $\lambda_{mf}$. Upper inset: Lyapunov spectrum for $N=100$. The dashed line denotes the mean field LE that can be estimate by using Eqs. (\ref{eq6}) and (\ref{eq7}) for the synchronized population.
}
\label{fig.spettro}
\end{figure}

The presence of alternating erratic and laminar phases makes difficult to employ 
the usual maximal Lyapunov Exponent to characterize the dynamics of the ICCs,
therefore to analyse the stability of this regime we prefer to determine the distribution 
$P(\Lambda)$ of the FTLE $\Lambda$. As a first result, we can affirm
that a non chaotic behaviour is associated to the laminar phases,
in fact during these phases $\Lambda$ is almost zero as shown Fig.~\ref{fig.Frequenze}(b). 
This reflects in the appearence of a peak around $\Lambda =0$ in the distribution $P(\Lambda)$,
as observable in the inset of Fig.~\ref{fig.spettro} (a).
In more details, the laminar phases are characterized by the synchronization of most 
part of the oscillators belonging to the chaotic population, which get entrained to the 
synchronized population. This can be understood looking at Fig. \ref{fig.Frequenze}(c), 
where the frequencies of two typical oscillators, one belonging to the chaotic and the
other to the synchronized population, are reported during chaotic and laminar phases. 
These two oscillators (as the majority of the oscillators) synchronize 
during the laminar phases, as observable by the fact that their frequencies
approach the constant average value associated to the fully synchronized regime, 
namely $\bar{\omega}=1-\frac{1}{2} \sin \gamma$. However, a few oscillators of the chaotic population 
do not get synchronized to all the others during the laminar phases. Instead,
they start oscillating with an unique frequency but with incoherent phases, giving rise to 
the oscillations observable in the order parameter and, thus, contributing to 
desynchronize the system. When the laminar phase ends, all the oscillators
in the chaotic population return to behave irregularly.

To characterize the degree of chaoticity of the erratic phase, we should give an 
estimate of the average Lyapunov exponent $\Lambda^{(*)}$ restricted to this phase. 
We proceeded as explained in Sect. II B, and as previously done in \cite{olmi2015}, 
in particular $\Lambda^{(*)}$ represents the maximum value of the probability distribution 
function $P(\Lambda)$ (shown in the inset of Fig.~\ref{fig.spettro}(a)),
once removed the peak around $\Lambda=0$ characteristic of the laminar phase.
It is interesting to examine the dependence of $\Lambda^{(*)}$ on the system
size. In particular, one can observe a clear decrease of $\Lambda^{(*)}$
with the system size, namely as $1/\ln(N)$ as clearly shown in  Fig.~\ref{fig.spettro}(a). 
Furthermore, from these data one can extrapolate the value of $\Lambda^{(*)}$ 
in the thermodynamic limit, which is definitely positive,
namely $\Lambda^{(*)} \simeq 0.022$. Therefore, we can affirm that the ICCs
remain chaotic for coupled rotators even for $N \to \infty$,
at variance with the weak chaotic regime observed for the Kuramoto model in~\cite{wolfrum2011spectral,popovych2005phase}. The logarithmic dependence of the maximal LE $\lambda_M$ with the system size has been previously found for globally coupled dissipative systems in~\cite{takeuchi2011}, where, 
the authors have shown analytically that
\begin{equation}
\label{eqdissipativesystem} 
\lambda_M(N) = \lambda_{mf} + \frac{D}{2} + \frac{a}{\ln(N)} + {\cal O}\left(\frac{1}{\ln^2 (N)}\right)
\enskip ;
\end{equation}
where $\lambda_{mf}$ is the mean field LE obtained by considering an isolated unit of 
the chaotic population forced by the fields $\rho^{(\sigma)}$, $D$ is the diffusion coefficient associated to the fluctuations of $\lambda_{mf}$.
As explained in Appendix A, these quantities can be numerically estimated giving
$\lambda_{mf} \simeq 0.0116(5)$ and $D \simeq 0.0180(10)$, therefore
in our case the expected asymptotic Lyapunov 
exponent should be $\lambda_M(\infty) \simeq 0.021(1)$, which is in good agreement with 
the previously reported numerical extrapolation, as shown in Fig.~\ref{fig.spettro}(a). 

A more detailed analysis of the stability of the chaotic phase can be attained by estimating the Lyapunov spectra for different sizes. The spectrum (see upper inset in Fig. ~\ref{fig.spettro}(b)) is composed by a positive part made of $N-2$ exponents and a negative part composed of $N$ exponents. Moreover two exponents are exactly zero: 
one is always present for systems with continuous time, while the second arises due to the invariance of 
Eq.~(\ref{eq1}) for uniform phase shifts.
In particular the negative part of the spectrum is composed by $N-1$ identical exponents, which measure the 
transverse stability of the synchronous solution, plus an isolated LE, which expresses the 
longitudinal stability of the synchronized population (as explained in Appendix B). The transverse stability
can be obtained by estimating the mean field LE associated to the synchronized family; the agreement with the numerical
data is perfect as shown in the inset of Fig. \ref{fig.spettro}(b) (for more details see Appendix A).
Furthermore, the central part of the positive spectrum tends to flatten towards the mean field
value $\lambda_{mf}$ for increasing system sizes (see Fig.~\ref{fig.spettro}(b) for N =100; 200; 400 and 800), 
while the largest and smallest positive LEs tend to split, in the same limit, from the rest
of the spectrum. This scaling of the Lyapunov spectra has been found to be a general property of fully coupled
dynamical systems in ~\cite{takeuchi2011,ginelli2011}, where it has been shown that the 
spectrum becomes asymptotically flat (thus trivially extensive) in the thermodynamic limit, 
but this extensive part is squeezed between the largest and smallest LEs, which
constitute two subextensive bands of order ${\cal O} (\log N)$. 
This scaling behavior, typical of fully coupled system, is in contrast with the results reported for chaotic 
chimeras in ~\cite{wolfrum2011,wolfrum2011spectral}, where the ring geometry strongly influences the stability
properties, thus indicating the dependence of the results for chaotic chimeras on the underlying network topology.

\subsection{Life-time of ICCs}

By measuring the life time of chaotic chimeras we have found that ICCs are transient states; in particular we observed for different inertia that chaotic chimeras converge to a regular (non-chaotic) state after a transient time $\tau$. In particular we have observed that ICCs decay either to a fully synchronized state or to a quasiperiodic chimera.

\begin{figure}
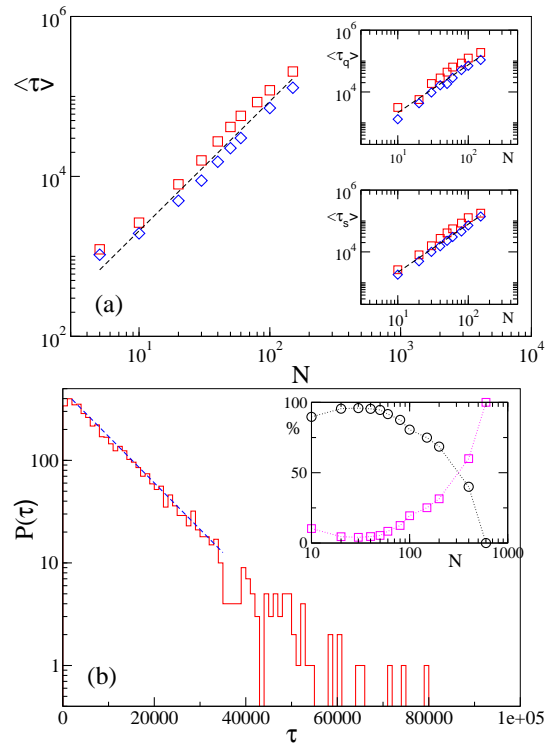

\includegraphics*[angle=0,width=7cm]{f11a.eps}
\includegraphics*[angle=0,width=7cm]{f11b.eps}
\caption{(a) Average life-time $<\tau>$ of ICCs vs $N$ for $m=10$ (red squares) and $m=8$ (blue diamonds). 
The black dashed line refers to a power law with exponent 1.60. 
In the insets are shown, for the same $m$ values, the partial results for the chaotic chimeras
that decay in a stable quasiperiodic chimera (upper inset) or in a fully synchronized state (lower inset). The black dashed line in the upper (lower) inset refers to a power law with exponent 1.55 (1.56). Data are averaged over 200-4000 different realizations of BSCs.
(b) Distribution $P(\tau)$ of the life-times $\tau$ of ICCs for $N=30$ and $m=6$ obtained 
from 4,000 data. The dashed blue curve is an exponential fit with a decay time $\tau^*=9,590(5)$.
Inset: percentage of states decaying towards a synchronized state (black circles) or towards a quasiperiodic chimera (magenta squares) as a function of the systems size.
}
\label{fig.transient1}
\end{figure}
By measuring the average life times $<\tau>$ for two inertia values, namely $m=8$ and 10, and 
various system sizes $5 \le N \le 150$, we have found a power-law divergence of the decaying
time with an exponent $\alpha \simeq 1.60(5)$ for $N \ge 10$  (see Fig.~\ref{fig.transient1}(a)).
Furthermore the power law divergence is still present if we consider separately the average life times 
$<\tau_q>$ ($<\tau_s>$) of ICCs decaying in a quasiperiodic chimera (synchronized state), 
as shown in the upper (lower) inset in Fig.~\ref{fig.transient1}(a).
Even though we were unable to verify that this scaling is present over more than one decade, due to 
computational problems, we can safely affirm that these times are not diverging exponentially with the size as reported 
in~\cite{wolfrum2011}. This is a further indication that the topology presently considered deeply influences the 
nature of the observed phenomenon.

\begin{figure} 
\includegraphics*[angle=0,width=7cm]{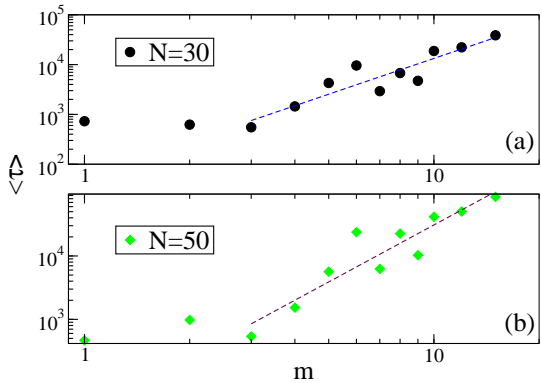}
\caption{Average life time $\langle \tau \rangle$ vs inertia for $N= 30$ (a) and 50 (b). 
The blue (maroon) dashed line in panel (a) (panel (b)) refers to a power law with exponent 2.38(5) (2.98(5)). The values of $\langle \tau \rangle$ are averaged over 3000 different realizations of BSCs.}
\label{fig.transient3}
\end{figure}

The life time distribution $P(\tau)$, obtained for different initial conditions, is exponential: 
as shown in Fig.~\ref{fig.transient1}(b) for $N=30$ and $m=6$.
The exponential decay can be fitted as $exp(-t/\tau^*)$, where the decay time $\tau^* \simeq 9,590$ is 
consistent with average life time value obtained for the same size and inertia, namely $\langle \tau \rangle> \simeq 9,975$. Thus indicating that this can be described as a Poisson process.
Moreover chaotic chimeras can preferentially decay to the synchronized state or to a broken symmetry state depending on the system size. In particular, for small sizes the decay to a synchronized state is preferred, while for larger sizes, ICCs decay more probably to a quasiperiodic chimera (as shown in the inset of Fig. ~\ref{fig.transient1}(b)).
Finally, we have tested the dependence of $\langle \tau \rangle$ on the inertia, for two system sizes, namely $N=30$ and 50 and we have observed that $\langle \tau \rangle$ is diverging as a power law with $m$ with  exponents $\simeq 2 -3$, as reported in Fig.~\ref{fig.transient3}. 

\section{Chaotic Two Populations}
\label{C2P}

In this Section the microscopic dynamics of stationary states emerging with UCs is characterized. 
For high inertia values both populations are usually chaotic but their dynamics
takes place on different macroscopic chaotic attractors, thus giving rise to broken symmetry
C2P states. Two examples are reported in in Figs.~\ref{fig.statictp} (a) and (b).
However, at variance with ICCs, C2Ps are not intermittent neither transient.

\begin{figure}[h]
\includegraphics*[angle=0,width=7cm]{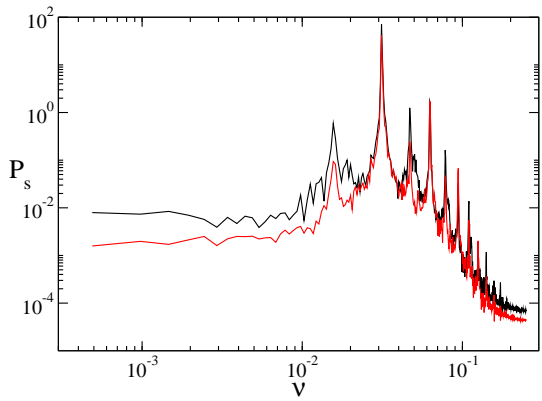}
\caption{ Power spectra $P_s$ as a function of $\nu$ for the two order 
parameters associated to the C2Ps state reported in Fig. \ref{fig.statictp} (a).
The  spectra $P_s$ are obtained by following each realization for a time span $T_s=100,000$ and by averaging over 
100 different initial conditions. Data refer to $m=10$, $N=200$.}
\label{fig.powerspectrum2}
\end{figure} 
 
To better characterize the C2Ps state, we have estimated the power 
spectra $P_s (\nu)$ associated with the order parameters $R^{(1)}$ and $R^{(2)}$
for the state shown in Fig. \ref{fig.statictp} (a).
These spectra, reported in Fig.~\ref{fig.powerspectrum2}, Fig. \ref{fig.statictp} (a)
show an almost flat behavior at low frequencies, thus being strongly different
from the spectrum of the ICCs shown in Fig.~\ref{fig.powerspectrumCC}. 
In particular, each spectrum resembles a Lorentzian with several
subsidiary peaks associated to harmonics and subharmonics of a
fundamental frequency $\nu_0 \simeq 0.03125$ ($T_0 = 1/\nu_0 = 32$).
The main peak, corresponding to $T_0$, is related to the mean period of the 
oscillations observable in the order parameters.

 \begin{figure}[h]
\includegraphics*[angle=0,width=9cm]{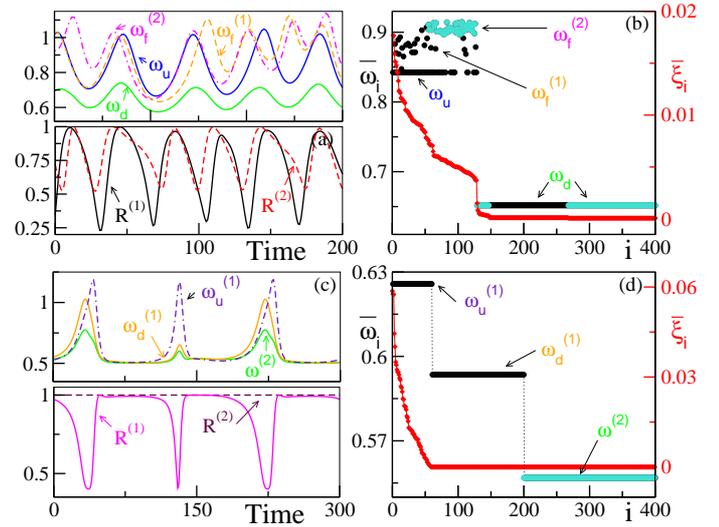}
\caption{Frequencies $\omega^{(1)}$ ($\omega^{(2)}$) of oscillators belonging to population 
$\sigma=1$ ($\sigma=2$) as a function of time, and the corresponding order parameter $R^{(1)}(t)$ ($R^{(2)}(t)$) 
for $m=9$ (a) and $m=3$ (c). Average components of the vector  $\left\lbrace \bar{\xi}_i\right\rbrace$  (red diamonds) 
together with the corresponding average frequencies of the oscillators $\bar{\omega}_i$ for population 
one (black circles) and two (turquoise circles), for $m=9$ (b) and $m=3$ (d). 
The indeces of the rotators are ordered accordingly to the value of $\bar{\xi}_i$. 
For more details on the hightlighted frequencies see the text.
Data in panels (a) and (c) are referred to the state shown in Fig. \ref{fig.statictp} (b) and
in panels (b) and (d) to the one reported in Fig. \ref{fig.statictp} (c).
The time averages have been performed over a time interval $T_s\simeq 5\times 10^4$.}
\label{fig.FrequenzeC2P}
\end{figure}

Moreover, the C2P states are characterized by different amplitude evolution of $R^{(1)}$ and $R^{(2)}$, 
being the differences due to the different number of isolated oscillators present in the two populations.  
In fact, oscillators of both populations share at least one common cluster,
characterized by a common average frequency, and the erratic dynamics is induced by the evolution of non 
clustered oscillators belonging to both populations. Let us examine for example 
the C2P state reported in Fig. \ref{fig.statictp} (b) for $m=9$. As shown in Fig.~\ref{fig.FrequenzeC2P}(b) 
the oscillators split in a common cluster plus a second cluster belonging to one population only and few non 
synchronized oscillators. The frequencies of the oscillators of the two populations belonging to the lower 
cluster $\omega_{d}$ are identical, while the frequencies belonging to the upper cluster $\omega_{u}$
are locked to those of the lower cluster; thus the two order parameters show similar oscillation periods as 
reported in Fig.~\ref{fig.FrequenzeC2P}(a). However, the oscillators belonging to the upper cluster and
few oscillators out of the clusters $\omega^{(\sigma)}_f$ ensure the microscopic dynamics to be chaotic, as can  
be proved by estimating the time averaged contribution $\bar{\xi}_i$, of the $i-$th oscillator, 
to the modulus of the maximal Lyapunov vector. In particular, as shown in Fig. ~\ref{fig.FrequenzeC2P}(b) 
the contribution  $\bar{\xi}_i$ of the oscillators belonging to the common cluster is essentially negligible
and the only relevant contributions comes from the oscillators with higher, non common, average frequencies.
These states resemble {\it imperfect chimeras} recently observed in experiments 
on coupled metronomes~\cite{kapitaniak2014} and in simulations of chains of Kuramoto oscillators with 
inertia~\cite{jaros2015}.

For low inertia values many states arise with different level of chaoticity. In Fig. ~\ref{fig.FrequenzeC2P}(c) and (d) the case shown in Fig.~\ref{fig.statictp} (c)
for $m=3$ is analyzed. In this example, one has a synchronized population and a 
second population with an almost regular behaviour, as discussed previously in Sect. IV.
Let us try to understand the origing of the residual chaoticity in the second population,
the oscillators belonging to the synchronized population share the same average frequency $\omega^{(2)}$, while the other population breaks up into two 
clusters: a lower one at frequency $\omega^{(1)}_d$ and an upper one at 
larger frequency $\omega^{(1)}_u$. As shown in Fig. ~\ref{fig.FrequenzeC2P}(d)
the only relevant contribution to maximal Lyapunov vector is due to the oscillator with higher
frequencies, as in the previous case, i.e. to the one belonging to the upper cluster.

\section{Conclusions}

In this work, we have characterized in details the dynamical properties of different symmetric or symmetry broken states emerging in a simple numerical model of heterogeneously coupled oscillators with inertia. The presence of inertia is a distinctive ingredient to observe the emergence of chaotic regimes. While for small inertia values, quasi-periodic (or in rare cases, breathing) chimeras coexist with the fully synchronized state, at large inertia values, two types of chaotic solutions are found depending on the initial conditions. For uniform initial conditions Chaotic Two Population states are observable, where the erratic dynamics is induced by the evolution of the non clustered oscillators belonging to both populations. On the other hand, both for initial conditions with a broken
symmetry  as well as for uniform initial conditions, Intermittent Chaotic Chimeras can emerge.
For ICCs the chaotic population exhibits turbulent phases interrupted by laminar regimes. 
While C2Ps are stable solutions, ICCs are transient states and their life-times diverge as a power-law with the size and the inertia value. Therefore, in the thermodynamic limit these states survive for infinite time. Moreover the Lyapunov analyses reveal chaotic properties in quantitative agreement with theoretical predictions for globally coupled dissipative systems~\cite{takeuchi2011}.

Chimera states have been initially observed in spatially extended systems with long-range coupling and, so far, this configuration has been considered analogous to its limit case represented by two globally coupled sub-populations with heterogeneous coupling ~\cite{abrams2008}. We clearly demonstrate that chaotic chimeras observed in spatially extended systems in ~\cite{wolfrum2011,wolfrum2011spectral} have completely different dynamical properties with respect to fully coupled sub-populations. This means that the relevance of the network topology for the stability properties of the chimera states, in particular of chaotic ones, has been so far overlooked and it suggests that also regular chimeras could reveal different stability properties related to the underlying topology. 

Finally, due to the introduction of inertia, stationary chimera states characterized by a fully synchronized population and a partially synchronized one, are no longer observable. In particular, performing a numerical continuation of the zero-inertia solution, we have shown that stationary chimeras cannot be continued to finite inertia. Instead, the continuation procedure give rises to 
breathing or quasi-periodic chimeras. This is probably due to the fact that the introduction of 
inertia corresponds to the addition of a new degree of freedom, thus giving rise to a singular perturbation of the dynamics.

\begin{acknowledgments}
We thank E. A. Martens, A. Politi, M. Rosenblum and A. Torcini for useful discussions and suggestions.
We acknowledge partial financial support from the Italian Ministry of University and Research within the 
project CRISIS LAB PNR 2011-2013. This work is part of the activity of the Marie Curie Initial Training 
Network 'NETT' project \# 289146 financed by the European Commission.
\end{acknowledgments}
 
\appendix

\section{Finite Size Scaling of the Maximal Lyapunov Exponent}

The mean field evolution of an oscillator $\phi^{(\sigma)}$ belonging to the population 
$\sigma$ for the considered model can be obtained by assuming that the influence 
of the oscillator on the network dynamics is negligible, while its dynamics is 
driven by the fields $\rho^{(1)}$ and $\rho^{(2)}$, associated to the two populations,
which are treated as external forcing fields. 
In particular, the mean field dynamics for an oscillator of population $\sigma$ can be written as
\begin{eqnarray}
\nonumber
&& m\ddot{\phi}^{(\sigma)} + \dot{\phi}^{(\sigma)}=\Omega
+\sum_{\sigma'=1}^2 K_{\sigma\sigma'}\left\lbrace Im\left[\rho^{(\sigma')}\right] \cos(\phi^{(\sigma)}+\gamma) \right.
\\ \label{eq6}  && \left.
- Re \left[\rho^{(\sigma')}\right] \sin(\phi^{(\sigma)}+\gamma)\right\rbrace \enskip .
\end{eqnarray}

If we linearize the previous equation we can obtain the evolution of an infinitesimal perturbations in the tangent space
\begin{eqnarray}
\nonumber &&
m\delta\ddot{\phi}^{(\sigma)} + \delta\dot{\phi}^{(\sigma)}=
-\sum_{\sigma'=1}^2 K_{\sigma\sigma'}\left\lbrace Im \left[\rho^{(\sigma')}\right] \sin(\phi^{(\sigma)}+\gamma) \right.
\\  \label{eq7} && \left.
+ Re \left[\rho^{(\sigma')}\right]\cos(\phi^{(\sigma)}+\gamma)\right\rbrace\delta\phi^{(\sigma)} \enskip.
\end{eqnarray}

Furthermore, using Equations (\ref{eq6}) and (\ref{eq7}), we have estimated numerically 
the associated maximal LE by applying the standard method reported in \cite{benettin1980}.
For each population $\sigma$ it is possible to find a maximal mean field LE
$\lambda_{0}^{(\sigma)}$. For the numerical integrations of Eqs (\ref{eq6}) and (\ref{eq7}) 
we have employed the fields  $\rho^{(1)}(t)$ and $\rho^{(2)}(t)$ obtained by simulating 
at the same time a two population network made of $2 N$ oscillators. In particular, for the ICC state where
$\sigma=1$ ($\sigma=2$) is the chaotic (synchronized) population,
$\lambda_{0}^{(1)}$ ($\lambda_{0}^{(2)}$) is positive (negative);
the mean field exponent $\lambda_{0}^{(2)}$ corresponds to
the transverse LE for the synchronized population and it is analogous to
the one discussed in Appendix B in the case of full synchronization of both populations (see Fig. \ref{fig.1}). 
While $\lambda_{0}^{(1)}$ is the value to which the most part of the positive LEs, associated to the 
chaotic population, tends for increasing system sizes, and it has been indicated in Sec. \ref{chaoticchimeras} as
$\lambda_{mf}$ (see Fig. \ref{fig.spettro} (b)).
 
For fully coupled dissipative systems, made of a single population of chaotic
units,  it has been demonstrated in~\cite{takeuchi2011} that 
in the thermodynamic limit the most part of the Lyapunov spectrum becomes flat 
assuming the value of the mean field LE for the considered system. 
Only a few LEs ${\cal O} (\ln N)$, locate at the extrema of the spectrum,
corresponding to the largest (smallest) values, exhibit different asymptotic
values. In particular, it has been shown that the maximal LE scale as
\begin{equation}
\label{eq8} 
\lambda_M = \lambda_{mf} + \frac{D}{2} + \frac{a}{\ln(N)} + {\cal O}\left(\frac{1}{\ln^2 (N)}\right)
\enskip ,
\end{equation}
where $\lambda_{mf}$ is the mean field LEs, and $D$ is the diffusion coefficient  associated to the 
fluctuations of the instantaneous mean field LE. 

In particular, $D$ takes into account the effect of the coupling with the other oscillators 
neglected in the estimation of the mean field LE; in order to estimate $D$ one has to measure 
the mean square displacement of the following quantity $[\ln d(t) - \lambda_{mf} t]$,
where  $d(t)=\sqrt{|\delta\dot{\phi}^{(1)}(t)|^2 + |\delta{\phi}^{(1)}(t)|^2}$
is the modulus of the infinitesimal vector appearing in Eq.~(\ref{eq7}).
Therefore for sufficiently long times one expects to observe the following scaling
\begin{equation}
\label{eq9} 
[\ln d(t) - \lambda_{mf} t]^2 = D t ,
\end{equation}
where $d(0)=1$.
In \cite{olmi2015} we have shown that  the
scaling reported in Eq. (\ref{eq8}) is optimally reproduced by the maximal LE measured
for our system composed of two populations and presenting a broken symmetry. Furthermore,
whenever we substitute $\lambda_{mf}$ with $\lambda_0^{(1)}$ in
Eq. (\ref{eq8}) and Eq. (\ref{eq9}), the Eq. (\ref{eq8})
gives also a very good estimate of the value attained by the maximal LE in the
thermodynamic limit, as shown in \cite{olmi2015}. These results confirm the validity of the theoretical estimate
reported in~\cite{takeuchi2011} also for system composed by more than one population,
and where the collective dynamics is represented by more than a single macroscopic field.

\section{Linear Stability of the Synchronized State}

\begin{figure}
\begin{center}
\includegraphics*[angle=0,width=7cm]{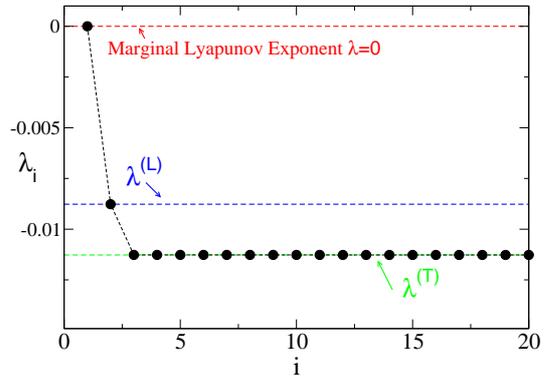}
\end{center}
\caption{Lyapunov spectrum $\lambda_i$ for the fully synchronized 
state. Analytical values: the marginal LE 
(dashed red line) corresponds to the maximal LE; the longitudinal 
LE $\lambda^{(L)}$ (dashed blue line) to the second exponent and 
the transverse LE $\lambda^{(T)}$ (dashed green line) to the last $2N-2$ degenerate exponents. 
The results of the numerical simulations are shown as black circles.
The data refer to $m=10$ and $N=10$.}
\label{fig.1}
\end{figure}

It is instructive to derive analytically the LEs for the fully
synchronized solution. Whenever the system is fully synchronized, Eq.~(\ref{eq2}) can be simplified by noticing 
that $\theta_i^{(1)}=\theta_j^{(2)}$ $\forall i,j$. Thus, Eq.~(\ref{eq2}) becomes
\begin{equation} 
\label{eq3bis}
 m \enskip\delta\ddot{\theta}_i^{(\sigma)} + \delta\dot{\theta}_i^{(\sigma)}= \sum_{\sigma'=1}^2 
\frac{K_{\sigma\sigma'}}{N} \cos(\gamma)\sum_{j=1}^N (\delta\theta_j^{(\sigma')}-\delta\theta_i^{(\sigma)}).
\end{equation}
Moreover it is useful to rewrite Eq.~(\ref{eq3bis}) as follows:
\begin{eqnarray} \nonumber
 && m \enskip\delta\ddot{\theta}_i^{(\sigma)} + \delta\dot{\theta}_i^{(\sigma)}= 
 -\frac{\cos(\gamma)}{2}\delta\theta_i^{(\sigma)} +  \frac{\mu\cos(\gamma)}{N} \delta\theta_i^{(\sigma)} 
 \\ \label{eq3} &+&
\frac{\mu}{N} \cos(\gamma)\sum_{j\neq i} \delta\theta_j^{(\sigma)}+
\frac{\nu}{N} \cos(\gamma)\sum_{j=1}^N \delta\theta_j^{(\sigma')}.
\end{eqnarray}
where now $\sigma\neq\sigma'$ is 1 or 2.

Starting from the previous equation it is possible to calculate both the longitudinal  
$\lambda^{(L)}$ and the transverse LE $\lambda^{(T)}$~\cite{pecora1995},
which are usually employed to characterize a synchronized system. In general,
even a chaotic state which displays a positive $\lambda^{(L)}$
can be fully synchronized, whenever all the transverse LEs are negative~\cite{pikovsky}. 
In the present case the shape of the Lyapunov spectrum $\{\lambda_i\}$, limited to the first half $i=1,\dots,2N$ 
is shown in Fig.~\ref{fig.1}. In the system under investigation one might expect two zero 
LE: one is always present for system with continuous time while the second zero LE 
is due to the invariance of the model under uniform phase shift. The figure reveals only one zero LE, 
instead of two as one might expect: the phase shift of all the phases corresponds to a perturbation along the 
orbit of the fully synchronized state, which explains why the two invariances, and thus LEs, coincide. 
Furthermore, one observes an isolated LE and a plateau of $2 N-2$ identical exponents. 

The first two exponents correspond to longitudinal exponents, while the $2N-2$ identical exponents correspond 
to transverse exponents.

The longitudinal exponents can be estimated by considering the average of all the perturbations
$Z^{(\sigma)}=\frac{1}{N}\sum_{j=1}^N \delta\theta_j^{(\sigma)}$ whose evolution
is ruled by 
\begin{equation}\label{eq4}
 m\ddot{Z}^{(\sigma)} + \dot{Z}^{(\sigma)}= \cos(\gamma)
\left[(\mu -\frac{1}{2})  Z^{(\sigma)} +  \nu Z^{(\sigma')} \right]
 \enskip ,
\end{equation}
and the associated eigenvalue equation reads as
\begin{eqnarray}
\nonumber
&& \left[m(\lambda^{(L)})^2 + \lambda^{(L)} \right] 
\left( \begin{array}{c} Z^{(1)} \\ Z^{(2)} \end{array}\right) = 
\\ \nonumber &&
\left( \begin{array}{cc} \cos(\gamma)(\mu-\frac{1}{2}) & \nu\cos(\gamma)
\\ \nu\cos(\gamma) & \cos(\gamma)(\mu-\frac{1}{2})  \end{array} \right) 
\left( \begin{array}{c} Z^{(1)} \\ Z^{(2)} \end{array} \right), 
\end{eqnarray}
yielding two second order secular equations 
\begin{equation}
\nonumber
m(\lambda^{(L)})^2 + \lambda^{(L)}=0 \enskip ,\qquad m(\lambda^{(L)})^2 + 
\lambda^{(L)} + 2 \cos(\gamma) \nu =0 \enskip .
\end{equation}
Each of the above equations admits two solutions that are symmetric with respect to $-1/2m$. 
The largest LEs are
\begin{equation}
\nonumber
 \lambda^{(L)} = 0 \enskip, \qquad 
  \lambda^{(L)} =\frac{-1 + \sqrt{1 - 8 m \nu \cos(\gamma) }}{2m}
  .
\end{equation}
The marginal LE, $\lambda^{(L)}=0$, is associated to a neutral
perturbation along the orbit, while the real part of the second longitudinal exponent 
is always negative. Therefore the achieved orbit is stable, provided that $m,\nu<\infty$ are finite and $\gamma \ne \pi/2$.

On the other hand, the transverse LE can be estimated by considering the evolution of the 
difference between two infinitesimal perturbations associated with two generic oscillators $i$ and $j$ of the 
same population, namely $W^{(\sigma)}=\delta\theta_j^{(\sigma)}-\delta\theta_i^{(\sigma)}$.
Using (\ref{eq3}), its temporal evolution is given by
\begin{equation}
\label{eq5}
 m\ddot{W}^{(\sigma)} + \dot{W}^{(\sigma)}=-\frac{\cos(\gamma)}{2}  W^{(\sigma)}.
\end{equation}
The associated secular equation reads
\begin{equation}
\nonumber
m (\lambda^2-\Omega^2)+ \lambda= -\frac{\cos(\gamma)}{2} \enskip,
\end{equation}
and is easily solved, yielding the maximal transverse LE
\begin{equation}
\nonumber
\lambda^{(T)} =\frac{-1 + \sqrt{1 - 2 m \cos(\gamma) }}{2m}
\end{equation}
whose real part is always negative for any finite inertia,
provided that $\gamma \ne \pi/2$.
 
The analytic predictions are in perfect agreement with the simulation results for a fully synchronized state, 
as shown in Fig.~\ref{fig.1}.  Furthermore, our analysis ensures that this state is always a stable solution for the 
system, given a finite inertia $m<\infty$, $A < 1$ and a phase lag $\gamma\neq \pi/2$. 

Whenever the phase lag is exactly
equal to $\pi/2$, $\lambda^{(L)}=\lambda^{(T)}=0$ and the system becomes highly degenerate, thus resembling the completely
integrable dynamics of phase oscillators with global cosine coupling studied in \cite{watanabe1993}.





\end{document}